\documentclass[aps,prl,reprint,superscriptaddress,nofootinbib]{revtex4-1}
\usepackage{graphicx}
\usepackage{dcolumn}
\usepackage{bm}
\usepackage{epstopdf}
\usepackage{amsmath}
\usepackage{hyperref}
\usepackage{color}
\usepackage{cancel}

\begin{document}



\title{Dibaryon with highest charm number near unitarity from lattice QCD}

\author{Yan Lyu
}
\email{Y. L. and H. T. contributed equally to this Letter and
should be considered as co-first authors. \\ helvetia@pku.edu.cn
}
\affiliation{State Key Laboratory of Nuclear Physics and Technology, School of Physics, Peking University, Beijing 100871, China }
\affiliation{Quantum Hadron Physics Laboratory, RIKEN Nishina Center, Wako 351-0198, Japan}
\author{Hui Tong
}
\email{Corresponding author.\\ tongh16@pku.edu.cn}
\affiliation{State Key Laboratory of Nuclear Physics and Technology, School of Physics, Peking University, Beijing 100871, China }
\affiliation{Interdisciplinary Theoretical and Mathematical Sciences Program (iTHEMS), RIKEN, Wako 351-0198, Japan}
\author{Takuya Sugiura
}
\affiliation{Interdisciplinary Theoretical and Mathematical Sciences Program (iTHEMS), RIKEN, Wako 351-0198, Japan}

\author{Sinya Aoki
}
\affiliation{Center for Gravitational Physics, Yukawa Institute for Theoretical Physics,
 Kyoto University, Kyoto 606-8502, Japan}
 \affiliation{Quantum Hadron Physics Laboratory, RIKEN Nishina Center, Wako 351-0198, Japan}
\author{\\ Takumi Doi
}
\affiliation{Quantum Hadron Physics Laboratory, RIKEN Nishina Center, Wako 351-0198, Japan}
\affiliation{Interdisciplinary Theoretical and Mathematical Sciences Program (iTHEMS), RIKEN, Wako 351-0198, Japan}
\author{Tetsuo Hatsuda
}
\affiliation{Interdisciplinary Theoretical and Mathematical Sciences Program (iTHEMS), RIKEN, Wako 351-0198, Japan}
\author{Jie Meng
}
\affiliation{State Key Laboratory of Nuclear Physics and Technology, School of Physics, Peking University, Beijing 100871, China }
\affiliation{Yukawa Institute for Theoretical Physics, Kyoto University, Kyoto 606-8502, Japan}
\author{Takaya Miyamoto
}
\affiliation{Quantum Hadron Physics Laboratory, RIKEN Nishina Center, Wako 351-0198, Japan}


\begin{abstract}
A pair of triply charmed baryons, $\Omega_{ccc}\Omega_{ccc}$,  is studied  as an ideal dibaryon system  by (2+1)-flavor lattice QCD with nearly physical light-quark masses and the relativistic heavy quark action with the physical charm quark mass.
The spatial baryon-baryon correlation is related to their scattering parameters  on the basis of  the HAL QCD method.
The $\Omega_{ccc}\Omega_{ccc}$ in the ${^1S_0}$ channel taking into account the Coulomb repulsion with the charge form factor of $\Omega_{ccc}$ leads to the scattering length $a^{\rm C}_0\simeq -19~\text{fm}$ and the effective range $r^{\rm C}_{\mathrm{eff}}\simeq 0.45~\text{fm}$.
The ratio $r^{\rm C}_{\mathrm{eff}}/a^{\rm C}_0 \simeq -0.024$, whose magnitude is considerably smaller than that of the dineutron ($-0.149$),  indicates that $\Omega_{ccc}\Omega_{ccc}$ is located in the unitary regime.
\end{abstract}



\maketitle




{\it Introduction.$-$} Quantum chromodynamics (QCD) is a fundamental theory of strong interaction and governs
 not only the interaction among quarks and gluons  but also the interaction between color-neutral hadrons.
In particular, the nucleon-nucleon ($NN$) interaction, which shows a characteristic mid-range attraction and a short-range repulsion,
  as well as the baryon-baryon ($BB$) interactions
are  important for  describing the nuclear structure and dense matter relevant to
  nuclear physics and astrophysics~\cite{Epelbaum2009,Meng2016,Shen2019,Drischler2019,Tong2020}.

Although the deuteron is the only stable bound state composed of two nucleons,
 there are possible bound or resonant dibaryons with and without strange quarks~\cite{Clement2017,Gal2015,Cho2017}.
 Among others,  $p\Omega(uudsss)$ ~\cite{Iritani2019} and $\Omega\Omega(ssssss)$~\cite{Gongyo2018},
  which were predicted  by lattice QCD (LQCD) simulations near the physical point~\cite{Aoki2020},
  stimulate  experimental searches in high energy hadron-hadron and heavy-ion collisions~\cite{Cho2017,Morita:2019rph,Acharya2020,Fabbietti2020}.

 As originally pointed out by Bjorken~\cite{Bjorken1985}, the triply charmed baryon (the charm number $C=3$) $\Omega_{ccc}$
  is stable against the strong interaction and  provides an ideal  ground to study the perturbative and non-perturbative aspects of QCD in the baryonic sector.
  Although it has not been observed yet  experimentally\footnote{Recently,
 excited states of $C=1$ baryon $\Omega_c$~\cite{Aaij2017Omega} and  a $C=2$ baryon  $\Xi_{cc}^{++}$~\cite{Aaij2017cascade}
  were   discovered at CERN LHC.}
   there have been numerous LQCD studies on its mass and electromagnetic form factor
   (see~\cite{Can2015} and references therein).
 Accordingly, it is timely to study the $\Omega_{ccc}\Omega_{ccc}$ as the simplest possible system to study heavy-baryon interactions.
Its recent phenomenological study using the constituent quark model can be found in Ref.~\cite{Huang2020}.

The purpose of this Letter is to  study  a system with the charm number $C=6$ system,
 $\Omega_{ccc}\Omega_{ccc}$ in the $^1 S_0$ channel,  for the first time  from first principle LQCD approach.
\footnote{In the charm number $C=3$ sector, there   exist  a few recent  studies on heavy dibaryons
    in  LQCD~\cite{Junnarkar2019} and  in the constituent quark model~\cite{Richard2020}.}
 The reason why we consider the $S$-wave and total spin $s=0$ system is that the Pauli exclusion between charm quarks
       at short distance does not operate in this channel, so that the maximum attraction is expected in comparison to
        other channels.  It is of critical importance  to examine  the scattering parameters such as the scattering length and
     the effective range to unravel  the properties of such heavy dibaryons near  threshold.
    The  HAL QCD method ~\cite{Ishii2007,Ishii2012,Aoki2020}, which treats the spatial correlation between two baryons on the lattice,
       provides a powerful tool for such analysis:   Indeed, we show below that   $\Omega^{++}_{ccc}\Omega^{++}_{ccc}$$(^1 S_0)$
        with both strong interaction and Coulomb repulsion
        is located near unitarity~\cite{Braaten2006,Cheng2010} just above the threshold  with a  large negative  scattering length.

{\it HAL QCD Method.$-$}
The crucial steps in the HAL QCD method~\cite{Ishii2007,Ishii2012,Aoki2020}  are to obtain the equal-time Nambu-Bethe-Salpeter (NBS) wave function $\psi(\bm{r})$ whose asymptotic behavior at a large distance reproduces the phase shifts, along with the corresponding two-baryon irreducible kernel
 $U(\bm{r},\bm{r}')$.   Since the same kernel $U(\bm{r},\bm{r}')$ governs all the elastic scattering states, separating the ground state and the excited states on the lattice, which is exponentially difficult for baryon-baryon interactions~\cite{Lepage1989, Iritani2019Jhep}, is not required  to calculate the physical observables~\cite{Ishii2012}.
The normalized four-point function (the $R$-correlator) related to the NBS wave function is defined as
\begin{equation}\label{equ1}
  \begin{split}
    R(\bm{r},t>0)&=\langle 0|\Omega_{ccc}(\bm{r},t)\Omega_{ccc}(\bm{0},t)\overline{\mathcal{J}}(0)|0\rangle/e^{-2m_{\Omega_{ccc}}t}\\
    &=\sum_n A_n \psi_n(\bm{r})e^{-(\Delta W_n)t}+O(e^{-(\Delta E^*)t}),
  \end{split}
\end{equation}
where $\Delta W_n =2\sqrt{m_{\Omega_{ccc}}^2+\bm{k}_n^2}-2m_{\Omega_{ccc}}$ with the baryon mass $m_{\Omega_{ccc}}$ and the relative momentum $\bm{k}_n$.
$O(e^{-(\Delta E^*)t})$ denotes the contributions from the inelastic scattering states with $\Delta E^*$ being the inelastic threshold, which are exponentially suppressed when $t \gg (\Delta E^*)^{-1}\sim \Lambda_{\mathrm{QCD}}^{-1}$ with $\Lambda_{\mathrm{QCD}}\sim $~300~MeV.
$\overline{\mathcal{J}}(0)$ is a source operator which creates two-baryon states with the charm number $C=6$ at Euclidean time $t=0$ and $A_n=\langle n | \overline{\mathcal{J}}(0) | 0\rangle $ with $|n\rangle$ representing the QCD eigenstates in a finite volume with $\Delta W_n< \Delta E^*$.
In this study, we take a local interpolating operator, $\Omega_{ccc}(x)\equiv \epsilon^{lmn}[c_l^T(x)\mathcal{C}\gamma_kc_m(x)]c_{n,\alpha}(x)$, where $l$, $m$, and $n$ stand for color indices, $\gamma_k$ being the Dirac
matrix, $\alpha$ being the spinor index, and $\mathcal{C} \equiv \gamma_4\gamma_2$ being the charge conjugation matrix.

When contributions from the inelastic scattering states are negligible ($t \gg (\Delta E^*)^{-1}$), the $R$-correlator satisfies~\cite{Ishii2012}
\begin{equation}\label{equ2}
  \left(\frac{1}{4m_{\Omega_{ccc}}}\frac{\partial^2}{\partial t^2}-\frac{\partial}{\partial t}-H_0\right) R(\bm{r},t) = \int d\bm{r}'U(\bm{r},\bm{r}')R(\bm{r}',t),
\end{equation}
where $H_0=-\nabla^2/m_{\Omega_{ccc}}$.
By using the derivative expansion at low energies, $U(\bm{r},\bm{r}')=V(r)\delta(\bm{r}-\bm{r}')+ \sum\limits_{n=1}V_{2n}(\bm{r})\nabla^{2n}\delta(\bm{r}-\bm{r}')$, the central potential $V(r)$ in the leading order (LO) is given as
\begin{equation}\label{equ3}
  V(r)=R^{-1}(\bm{r},t)\left(\frac{1}{4m_{\Omega_{ccc}}}\frac{\partial^2}{\partial t^2}-\frac{\partial}{\partial t}-H_0\right) R(\bm{r},t).
\end{equation}
The spatial and temporal derivatives of $R(\bm{r},t)$ on the lattice  are calculated in central difference scheme by using the nearest neighbor points.
To extract the total spin $s=0$, the following interpolating operators for the $\Omega_{ccc}\Omega_{ccc}$ system is adopted, $[\Omega_{ccc}\Omega_{ccc}]_0=\frac{1}{2}(\Omega_{ccc}^{3/2}\Omega_{ccc}^{-3/2}-\Omega_{ccc}^{1/2}\Omega_{ccc}^{-1/2}+\Omega_{ccc}^{-1/2}\Omega_{ccc}^{1/2}
-\Omega_{ccc}^{-3/2}\Omega_{ccc}^{3/2})$.
Here the spin and its $z$ component of the interpolating operator $\Omega^{s_z}_{ccc}$  are $3/2$ and $s_z=\pm 3/2,\pm 1/2$, respectively,
and $\Omega^{s_z}_{ccc}$ is constructed by spin projection as shown in Ref.~\cite{Yamada2015}.
To obtain the orbital angular momentum $L=0$  on the lattice, the projection to $A_1$ representation of the cubic group $SO(3,\mathbf{Z})$  is employed;  $P^{A_1}R(\bm{r},t)=\frac{1}{24}\sum\limits_{\mathcal{R}_i\in SO(3,\mathbf{Z})}R(\mathcal{R}_i[\bm{r}],t)$.
Note that  $V(r)$ in Eq.~\eqref{equ3} contains the channel coupling effect such as  $^1S_0$-$^5D_0$ mixing 
 and should be considered as an ``effective" potential projected onto the $S$-wave state \cite{Aoki2012}.

{\it Lattice setup.$-$}  (2+1)-flavor gauge configurations are generated on the $L^4 = 96^4$ lattice with the Iwasaki gauge action at $\beta=1.82$ and nonperturbatively $O(a)$-improved Wilson quark action combined with stout smearing at nearly physical quark masses ($m_\pi\simeq146~\text{MeV}$ and $m_K\simeq525~\text{MeV}$) \cite{Ishikawa2016}.
The lattice cutoff is $a^{-1}\simeq2.333~\text{GeV}$ ($a\simeq0.0846~\text{fm}$), corresponding to $La\simeq8.1~\text{fm}$, which is sufficiently large to accommodate two heavy baryons.
For the charm quark, we employ the relativistic heavy quark (RHQ) action in order to remove the leading order and the next-to-leading order cutoff errors
associated with the charm quark mass \cite{Aoki20013}.
We use two sets (set~$1$ and set~$2$) of RHQ parameters determined in Ref.~\cite{Namekawa2017} so as to interpolate the physical charm quark mass and reproduce the dispersion relation for the  spin-averaged $1S$ charmonium, i.e.
a weighted average of the  spin-singlet state $\eta_c$ and the spin-triplet  state $J/\Psi$.

For the source operator $\overline{\mathcal{J}}(0)$,   we use the wall type with the Coulomb gauge fixing. We employ the periodic (Drichlet) boundary condition for spatial (temporal) direction.
  We use 112 gauge configurations which are picked up one per ten trajectories.
   In order to reduce statistical fluctuations, forward and backward propagations are averaged,
    and four times measurements are performed by shifting source position along the temporal direction for each configuration.
  Then, the total measurements amount to $896$ for each set.
The statistical errors are estimated by the jackknife method with a bin size of $14$ configurations.
A comparison with a bin size of $7$ configurations shows that the bin size dependence is small.
The quark propagators are calculated by the Bridge++ code \cite{Bridge}, and the unified contraction algorithm is utilized to obtain the correlation functions~\cite{Doi2013}.

\begin{table}[htbp]
\caption{Spin-averaged $1S$ charmonium mass (($m_{\eta_c}+3m_{J/\Psi})/4$) and the  $\Omega_{ccc}$ mass ($m_{\Omega_{ccc}}$)
  calculated in  set $1$ and set $2$
  with the statistical errors.
The third row  shows the interpolated
 values  obtained from  set $1$ and set $2$.  Experimental value of $(m_{\eta_c}+3m_{J/\Psi})/4$ is shown in the last row.
}
\begin{tabular}{ccc}
  \hline\hline
   &~~~ $(m_{\eta_c}+3m_{J/\Psi})/4~\text{[MeV]}$  &~~~ $m_{\Omega_{ccc}}~\text{[MeV]}$ \\
  \hline
  set $1$        &~~~ $3096.6(0.3)$ &~~~ $4837.3(0.7)$ \\
  set $2$        &~~~ $3051.4(0.3)$ &~~~ $4770.2(0.7)$ \\
  Interpolation  &~~~ $3068.5(0.3)$ &~~~ $4795.6(0.7)$ \\
  Exp.           &~~~ $3068.5(0.1)$ &~~~-  \\
  \hline\hline
\end{tabular}
\label{tab1}
\end{table}

Masses for spin-averaged $1S$ charmonium ($(m_{\eta_c}+3m_{J/\Psi})/4$) and $\Omega_{ccc}$ baryon ($m_{\Omega_{ccc}}$) calculated in set $1$ and set $2$ by utilizing the single exponential fitting from the interval $t/a=25-35$ are listed in Table.~\ref{tab1}, together with
 the  values from linear interpolation ($0.3786\times\mathrm{set~1}+0.6214\times\mathrm{set~2}$) as well as the experimental value.
Our result for $m_{\Omega_{ccc}}$ is consistent with $4789(6)(21)$ MeV obtained by the (2+1)-flavor PACS-CS configurations~\cite{Namekawa2013}.
We have checked  that our results for hadron masses are unchanged within errors by  the fitting interval $t/a=30-35$.

\begin{figure}[t]
  \centering
  \includegraphics[width=8.7cm]{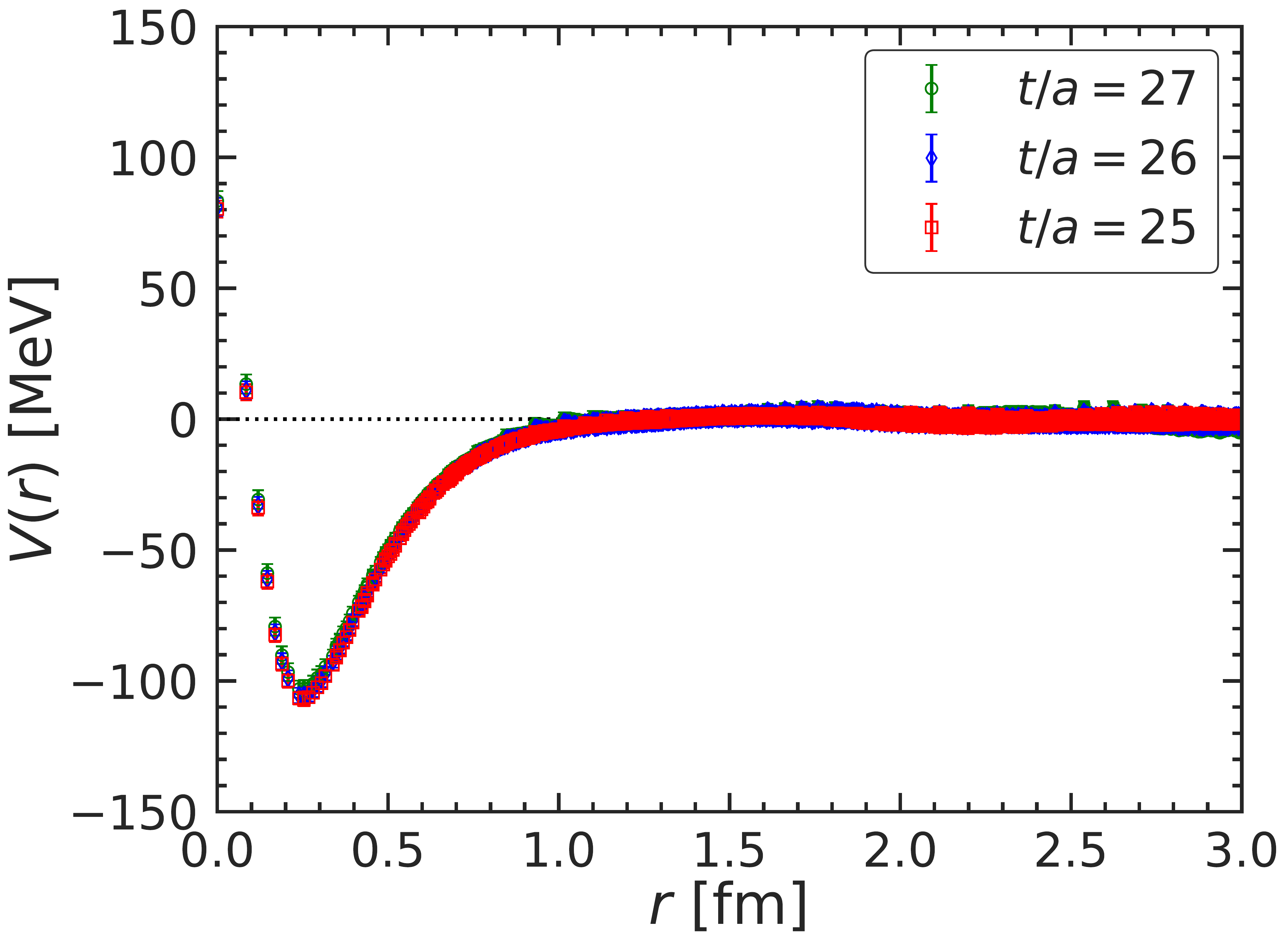}
  \caption{(Color online).
  The $\Omega_{ccc}\Omega_{ccc}$ potential $V(r)$ in the ${^1S_0}$ channel as a function of separation $r$ at Euclidean time $t/a= 25$ (red square), $26$ (blue diamond) and $27$ (green circle).
  }
  \label{Fig1}
\end{figure}

{\it Numerical results.$-$} The $\Omega_{ccc}\Omega_{ccc}$ potential $V(r)$ in the ${^1S_0}$ channel from the interpolation between set~1 and set~2  is shown in Fig.~\ref{Fig1} for $t/a= 25,~26$ and $27$ (see Supplemental Material~\cite{Supp} for the $t$-dependence of $V(r)$ in a wide range of $t$).
 Since the potentials from set 1 and set 2 are found to be consistent
  within statistical errors, the uncertainty in the interpolation is negligible.
Our choice $t/a = 26$ corresponds to $t\simeq2.2~\text{fm}$;  this is large enough in comparison to the typical
 length scale $\Lambda^{-1}_{\text{QCD}}\sim 0.7~\text{fm}$ characterizing the inelastic states, and is
  small enough to avoid  large statistical errors.
We find that the potentials for $t/a=$ 25, 26 and 27 are consistent with each other within statistical errors.
This  indicates that systematic errors due to inelastic states and  higher order terms of the derivative expansion do not largely exceed  the size  of statistical errors~\cite{Ishii2012}  as we show below.

We find that the potential $V(r)$ is repulsive at short-range and attractive at mid-range, which has the same qualitative behaviors with the $NN$ potential~\cite{Doi2017} and the $\Omega\Omega$ potential~\cite{Gongyo2018}.
The magnitude of the potential in the repulsive region $r < 0.25 $ fm (corresponding to $dV(r)/dr < 0$)
for $\Omega_{ccc}\Omega_{ccc}$  is an order of magnitude  smaller than that of $\Omega\Omega$ obtained by the same method~\cite{Gongyo2018}. This may be qualitatively
explained by the phenomenological quark model~\cite{Oka1987}   as
  the color-magnetic interaction between constituent quarks is proportional to the square of reciprocal constituent
  quark mass. Qualitatively, $V_{\rm cm}^{cc}/V_{\rm cm}^{ss} = (m_s^*/m_c^*)^2 \sim (500/1500)^2 \sim 0.1$, where $V_{\rm cm}^{ff'}$ is the color-magnetic interaction between the quarks with flavor $f$ and $f'$ with  $m_f^*$ being the constituent quark mass.
  On the other hand, the attraction in the region  $r > 0.25 $ fm (corresponding to $dV(r)/dr > 0$)
  may  originate from the exchange of charmed mesons  or rather  be attributed to the direct exchange of charm quarks and/or multiple gluons.
As can be seen in Fig.~\ref{Fig1}, the range of the potential is much smaller than the size of the lattice volume, indicating that the finite volume artifact is negligible.

In order to convert the potential to physical observables such as the scattering phase shifts and binding energy, we perform the uncorrelated fit for $V(r)$ in Fig.~\ref{Fig1} in the range $r\leq2.5$ fm by three-range Gaussians, $V_{\text{fit}}(r)=\sum\limits_{i=1,2,3} \alpha_i\exp(-\beta_i r^2)$.
Fitting parameters with $t/a=26$ for example  are $(\alpha_1, \alpha_2, \alpha_3)=(239.5(3.0), -62.7(50.8), -98.8(50.3))$ in MeV and $(\beta_1, \beta_2, \beta_3)=(48.5(1.4), 7.8(2.6), 3.4(0.8))$ in $\mathrm{fm}^{-2}$
with an accuracy of $\chi^2/\text{d.o.f.}\sim1.05$.

\begin{figure}[t]
  \centering
  \includegraphics[width=8.7cm]{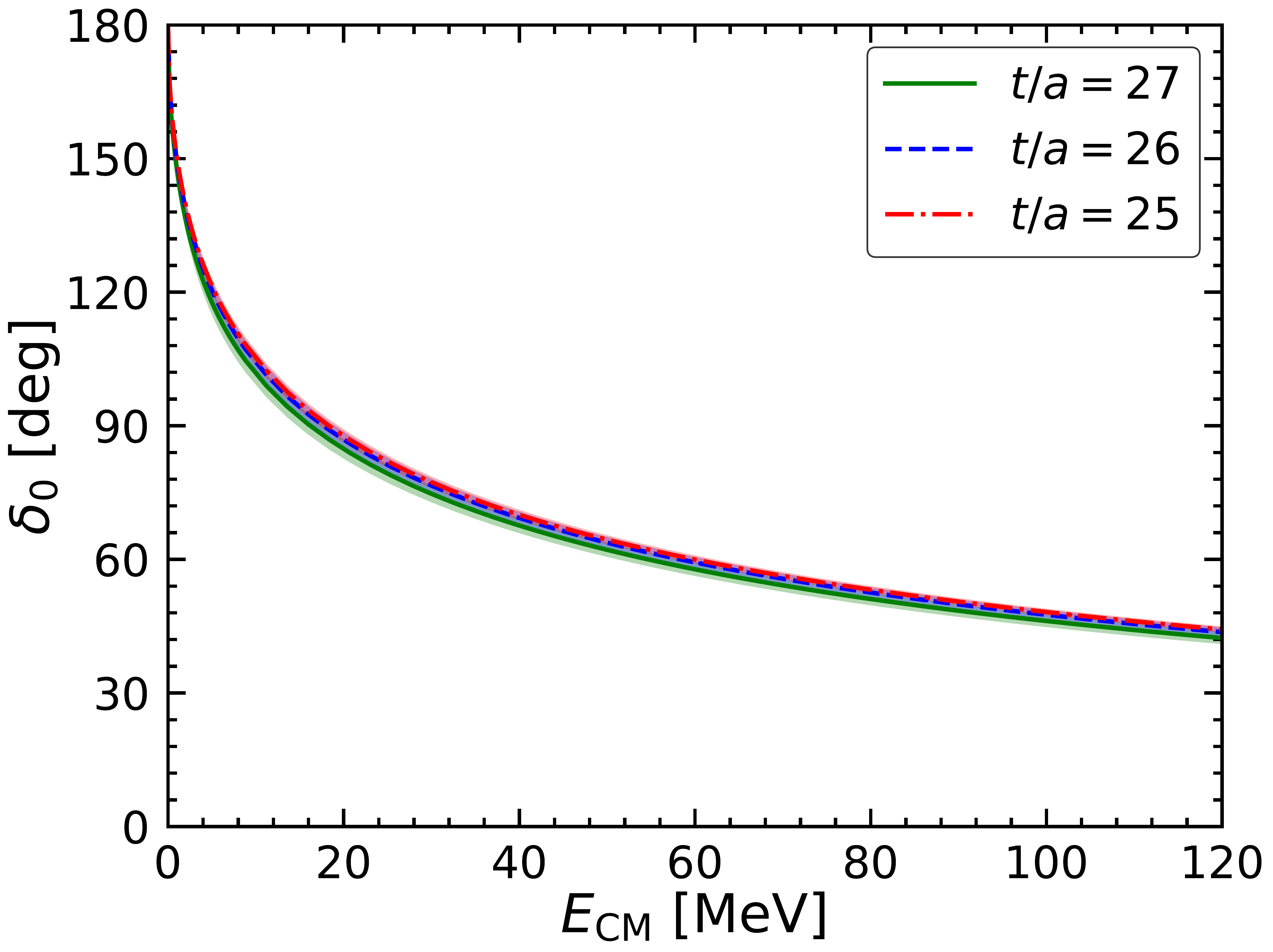}
  \caption{(Color online).
    The $\Omega_{ccc}\Omega_{ccc}$ scattering phase shifts $\delta_0$ in the $^1S_0$ channel obtained from the  potential $V(r)$ at $t/a=25, 26$, and $27$ as a function of the center of mass kinetic energy $E_{\mathrm{CM}}$.
  }
  \label{Fig2}
\end{figure}

In Fig.~\ref{Fig2}, we show the $\Omega_{ccc}\Omega_{ccc}$ scattering  phase shifts $\delta_0$ in the $^1S_0$ channel
 calculated by solving the Schr$\rm{\ddot{o}}$dinger equation with the potential $V(r)$ at $t/a = 25,~26$, and $27$.
The relativistic  kinetic energy is defined as $E_{\mathrm{CM}}=2\sqrt{k^2+m_{\Omega_{ccc}}^2}-2m_{\Omega_{ccc}}$ with a momentum $k$ in the center of mass frame. The error bands reflect the statistical uncertainty of $V(r)$.
In all three cases, the phase shifts start from $180^\circ$ at $E_{\mathrm{CM}}= 0$, which indicates the existence of a bound state in $\Omega_{ccc}\Omega_{ccc}$ system without Coulomb repulsion.

The low-energy scattering parameters are extracted by using the effective range expansion up to the next-to-leading order (NLO), $k\cot\delta_0 = -\frac{1}{a_0}+\frac{1}{2}r_{\mathrm{eff}}k^2+O(k^4)$,
where $a_0$ and $r_{\rm{eff}}$ are the scattering length and the effective range, respectively. The results  are
\begin{equation}\label{equ4}
  \begin{split}
   & a_0=1.57(0.08)(^{+0.12}_{-0.04})~\text{fm},\\
   & r_{\mathrm{eff}}=0.57(0.02)(^{+0.01}_{-0.00})~\text{fm}.
  \end{split}
\end{equation}
The central values and the statistical errors in the first parentheses are obtained at $t/a = 26$, while the systematic errors in the last parentheses are estimated from the values at $t/a=25$, $26$ and $27$, which originates from the inelastic states and the higher order terms of the derivative expansion.

\begin{figure}[t]
  \centering
  \includegraphics[width=9.0cm]{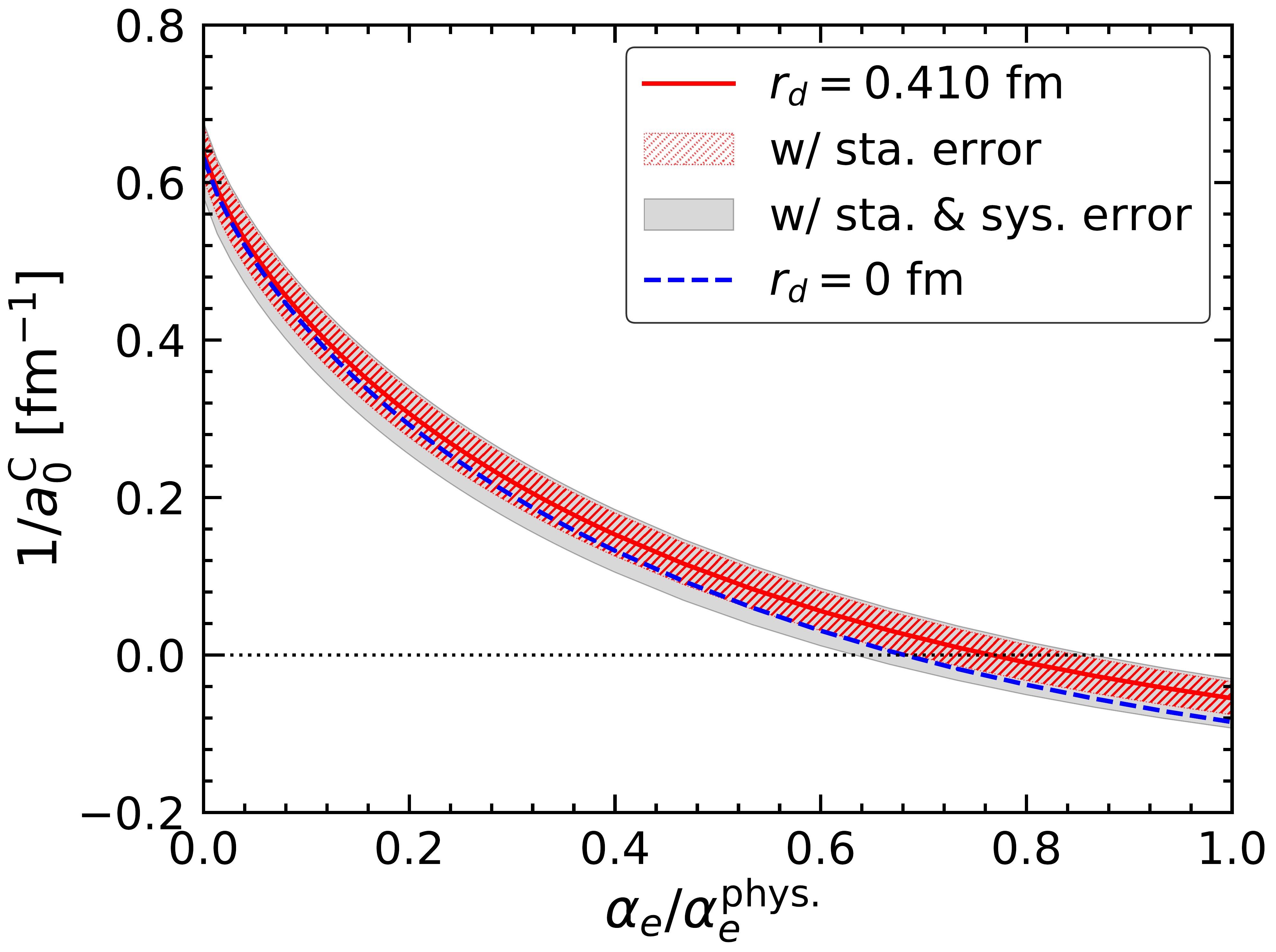}
  \caption{(Color online).
    The inverse of the scattering length $1/a^{\rm C}_0$ as a function of $\alpha_e/\alpha^{\mathrm{phys.}}_e$. The red solid line is the central values for $r_d=0.410$~fm.
The statistical errors are shown by the inner band (red), while the outer band (gray)
      corresponds to the statistical and systematic errors added in quadrature.
        The blue dashed line corresponds to the central values for $r_d=0$ fm.
  }
  \label{Fig3}
\end{figure}

The binding energy $B$ and the root-mean-square distance $\sqrt{\langle r^2\rangle}$ of  the bound $\Omega_{ccc}\Omega_{ccc}$
state are obtained from the potential $V(r)$ as
\begin{equation}\label{equ5}
 \begin{split}
  &B=5.68(0.77)(^{+0.46}_{-1.02})~\text{MeV},\\
  &\sqrt{\langle r^2\rangle}=1.13(0.06)(^{+0.08}_{-0.03})~\text{fm}.
 \end{split}
\end{equation}
These results
 are consistent with the general formula for loosely bound states \cite{Braaten2006,Naidon2017} with scattering parameters $a_0$ and $r_\text{eff}$: $B=1/(m_{\Omega_{ccc}}r^2_\text{eff})(1-\sqrt{1-(2r_\text{eff}/a_0)})^2\simeq 5.7~\text{MeV}$ and $\sqrt{\langle r^2\rangle}=a_0/\sqrt{2}\simeq 1.1~ \text{fm}$.

Since the binding energy and the size of the bound state from the strong interaction are
not large, we need to take into  account the
 Coulomb repulsion $V^{\text{Coulomb}}(r)$ between  $\Omega^{++}_{ccc}$s with finite spatial size.
 For this purpose,  we consider the dipole form factor for $\Omega_{ccc}^{++}$ according to  the LQCD study on
 the charge distribution of heavy baryons~\cite{Can2015}:
  In the coordinate space, it corresponds to an exponential charge distribution $\rho(r)=12\sqrt{6}/(\pi r_d^3)e^{-2\sqrt{6}r/r_d}$, where the charge radius $r_d=\sqrt{|\langle r^2\rangle_{\mathrm{charge}}|}$ of $\Omega_{ccc}^{++}$ is taken to be   $r_d=0.410(6)~\mathrm{fm}$~\cite{Can2015}.
Then, we have
\begin{equation}
V^{\text{Coulomb}}(r)= \alpha_e \iint d^3r_1 d^3r_2 \frac{\rho(r_1) \rho(|\vec{r}_2 - \vec{r}|)}{|\vec{r}_1-\vec{r}_2|}=\frac{4\alpha_e}{r}F(x),
\end{equation}
where $ x=2 \sqrt{6}r/r_d$ and $F(x) = 1 - e^{-x} (1 + \frac{11}{16} x + \frac{3}{16} x^2 + \frac{1}{48} x^3)$.
 The effective range expansion with Coulomb repulsion
 is written as
 \begin{equation}
 k\left[C^2_\eta\cot\delta^{\rm C}_0(k)+2\eta h(\eta)\right]=-\frac{1}{a^{\rm C}_0}+\frac{1}{2}r^{\rm C}_{\mathrm{eff}}k^2+O(k^4) ,
 \end{equation}
 where $\delta^{\rm C}_0(k)$ is the phase shift in the presence of Coulomb repulsion, $C^2_\eta=\frac{2\pi\eta}{e^{2\pi\eta}-1}$, $\eta={2\alpha_e m_{\Omega_{ccc}}}/{k}$, $h(\eta)=\text{Re}[\Psi(i\eta)]-\ln(\eta)$, and $\Psi$ is the digamma function~\cite{Burke2011}.
 To see the effect of the Coulomb repulsion, we vary $\alpha_e$ from zero to  the physical value $\alpha^{\mathrm{phys.}}_e=1/137.036$ below.
Note that the systematic errors originated from the uncertainty in $r_d$ are found to be much smaller than the statistical errors and are neglected.

In Fig.~\ref{Fig3}, we show the inverse of scattering length $1/a^{\rm C}_0$ under the change of $\alpha_e/\alpha^{\mathrm{phys.}}_e$  from 0 to 1.
 Due to the large cancellation between the attractive strong interaction and the Coulomb repulsion,
  the result at  $\alpha_e/\alpha^{\mathrm{phys.}}_e = 1$ is located very close to unitarity with a large scattering length
\begin{equation}\label{equ6}
  \begin{split}
   & a^{\rm C}_0=-19(7)(^{+7}_{-6})~\text{fm},\\
   & r^{\rm C}_{\mathrm{eff}}=0.45(0.01)(^{+0.01}_{-0.00})~\text{fm}.
  \end{split}
\end{equation}
The ratio $r^{\rm C}_{\rm eff}/a^{\rm C}_0=-0.024(0.010)(^{+0.006}_{-0.014})$ is considerably smaller in magnitude than that of the dineutron  ($-0.149$).

\begin{figure}[t]
  \centering
  \includegraphics[width=9.0cm]{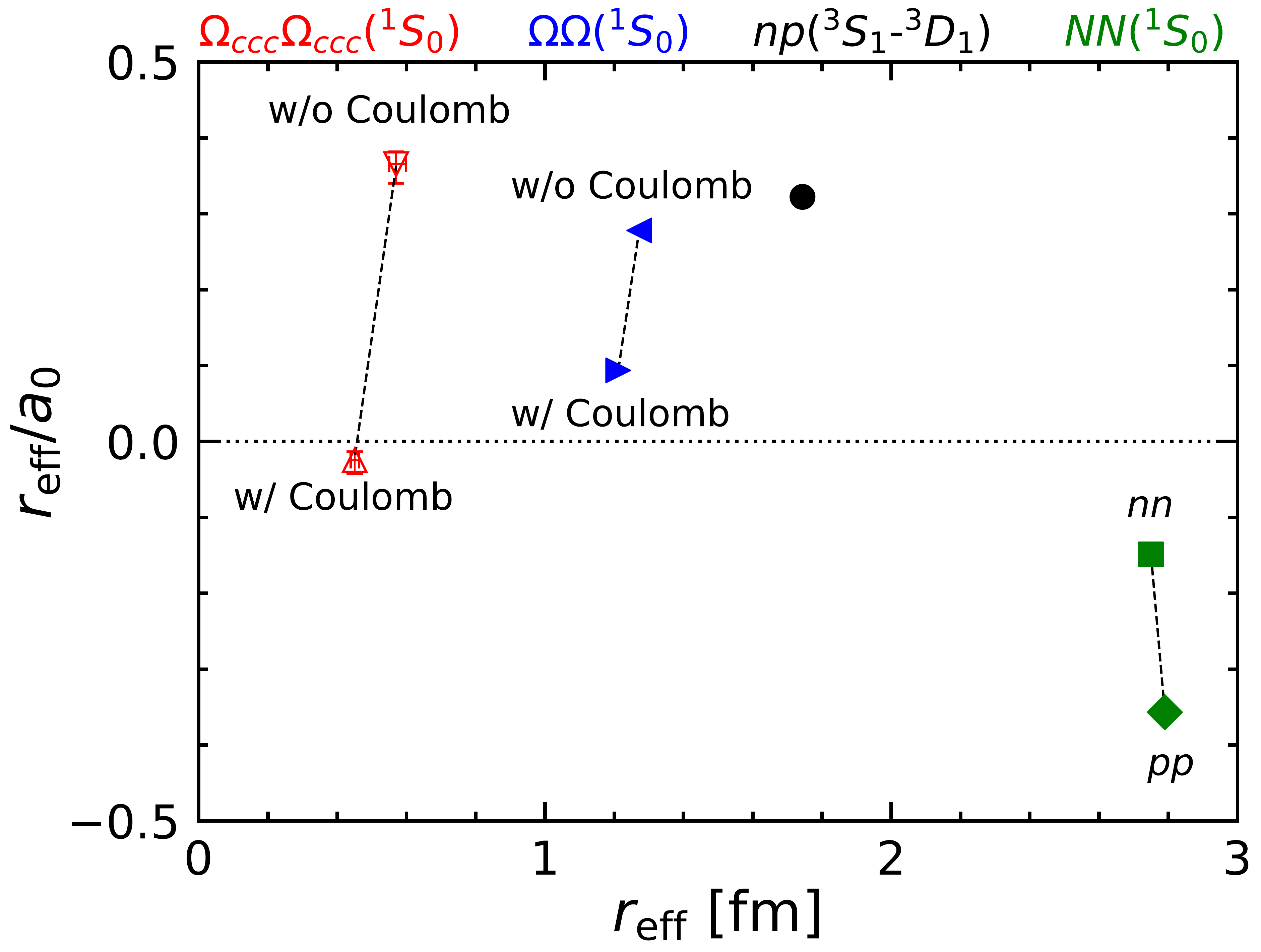}
  \caption{(Color online).
    The dimensionless ratio of the effective range $r_{\text{eff}}$ and the scattering length $a_0$ as a function of $r_{\text{eff}}$. The red up(down)-pointing triangle and the blue right(left)-pointing triangle correspond to $\Omega_{ccc}\Omega_{ccc}$ system and $\Omega\Omega$ system in the ${^1S_0}$ channel with(without) the Coulomb repulsion respectively. The black circle represents $NN$ system in the ${^3S_1}$-${^3D_1}$ channel. The green square~($nn$) and diamond~($pp$) correspond to $NN$ system in the ${^1S_0}$ channel.  The error bars for $\Omega_{ccc}\Omega_{ccc}$ are the quadrature of the statistical and systematic errors in Eqs.~\eqref{equ4} and \eqref{equ6}.
  }
  \label{Fig4}
\end{figure}

In Fig.~\ref{Fig4}, we plot  the dimensionless ratio $r_{\text{eff}}/a_0$ as a function of $r_{\text{eff}}$ for $\Omega^{++}_{ccc}\Omega^{++}_{ccc} ({^1S_0})$ and $\Omega^-\Omega^- ({^1S_0})$ with (without) Coulomb repulsion together with the experimental values for
 $NN$(${^3S_1}$-${^3D_1}$)~\cite{Hackenburg2006} and $NN({^1S_0})$~\cite{Bergervoet1988,Slaus1989}.
Note that we consider the Coulomb repulsion in $\Omega^-\Omega^- ({^1S_0})$ with the charge radius
 $r_d=0.57~\mathrm{fm}$ for $\Omega^-$~\cite{Can2015}.\footnote{In Ref.\cite{Gongyo2018}, $\Omega^-$ was assumed to be point-like
  in charge distribution, which
  overestimates the repulsion and increases the scattering length by $1$~fm.}
Among all those dibaryon systems, $\Omega^{++}_{ccc}\Omega^{++}_{ccc} ({^1S_0})$ is the closest to unitarity.
Note also that  the nearly unitary binding of both  $\Omega^{-}_{sss}\Omega^{-}_{sss}$($^1 S_0$) and  
$\Omega^{++}_{ccc}\Omega^{++}_{ccc}$($^1 S_0$) originates from a subtle cancellation among
 the potential energy, the kinetic energy and the Coulomb repulsion.

Finally, we briefly discuss other possible systematic errors in this work:
  (i) The finite cutoff effect is ${\cal O}(\alpha_s^2 a\Lambda_{\mathrm{QCD}}, (a\Lambda_{\mathrm{QCD}})^2)$
  thanks to the RHQ action for the charm quark and the non-perturbative $O(a)$ improvement for light ($u,d,s$) quarks,
  and thus amounts to be ${\cal O}(1) \%$.
  (ii) In the vacuum polarization,
  light quark masses are slightly heavier than the physical ones
  and charm quark loop is neglected.
  The former effect is expected to be small since light quarks are rather irrelevant for $\Omega_{ccc}\Omega_{ccc}$ system.
  In fact, the range of the $\Omega_{ccc}\Omega_{ccc}$ potential is found to be shorter than 1~fm.
  The latter effect is suppressed due to the heavy charm quark mass, and is typically ${\cal O}(1) \%$~\cite{FLAG2019}.
  These estimates for (i) and (ii) are also in line with the observation that
  our value of $m_{\Omega_{ccc}}$ is consistent with that in the literature or has deviation of $\sim 1\%$ at most,
  where we refer to LQCD studies by
  (2+1)-flavor at the physical point with finite $a$~\cite{Namekawa2013},
  (2+1)-flavor with chiral and continuum extrapolation~\cite{Brown2014} and
  (2+1+1)-flavor with chiral and continuum extrapolation~\cite{Briceno2012,Alexandrou2014}.
In the future, these systematic errors will be evaluated explicitly.
Moreover, a finite volume analysis with proper projection of the sink operator~\cite{Iritani2019Jhep} for the $\Omega_{ccc}\Omega_{ccc}$ system indicates that the truncation effect in the derivative expansion of $U(\bm r,\bm r')$ is small~\cite{Supp}. Further details will be reported elsewhere.

{\it Summary and discussions.$-$}  In this Letter, we presented a first investigation on the scattering properties of the
 $\Omega_{ccc}\Omega_{ccc}$  on the basis of the (2+1)-flavor lattice QCD simulations
  with physical charm mass and nearly physical light quark masses.
 The potential for $\Omega_{ccc}\Omega_{ccc}$($^1 S_0$)  obtained by the time-dependent HAL QCD method  without the Coulomb interaction
  shows a weak repulsion at short distance surrounded by a relatively strong attractive well, which  leads to a
   most charming ($C=6$)  dibaryon with  the binding energy  $B\simeq 5.7~\text{MeV}$ and the size
 $\sqrt{\langle r^2\rangle}\simeq 1.1 ~\text{fm}$.
  By taking into account the Coulomb repulsion between $\Omega^{++}_{ccc}$s
  with their charge form factor obtained from LQCD,
  the $\Omega^{++}_{ccc}\Omega^{++}_{ccc}$($^1 S_0$)  system turns into  the unitary region with  $r^{\rm C}_{\rm eff}/a^{\rm C}_0\simeq-0.024$.
  This provides good information toward the understanding of the interaction between heavy baryons. It is an interesting future work to study $\Omega^{-}_{bbb}\Omega^{-}_{bbb}$($^1 S_0$) for revealing the quark mass dependence of the scattering parameters.
   Finally, our results may further stimulate the future experimental activities to measure
    pair-momentum correlations of heavy baryons in high energy $pp$, $pA$ and $AA$ collisions~\cite{Cho2017,Fabbietti2020}.


\

\begin{acknowledgments}
{\it Acknowledgments.$-$}
We thank the members of HAL QCD Collaboration for technical supports and stimulating  discussions.
We thank Yusuke Namekawa for providing the RHQ parameters.
We thank members of PACS Collaboration for the gauge configuration generation conducted
on the K computer at RIKEN.
The lattice QCD measurements have been performed on HOKUSAI supercomputers at RIKEN.
This work was partially supported by HPCI System Research Project
(hp120281, hp130023, hp140209, hp150223, hp150262, hp160211, hp170230, hp170170, hp180117, hp190103).
We thank ILDG/JLDG~\cite{ldg}, which serves as an essential infrastructure in this study.
We thank the authors of cuLGT code~\cite{Schrock2013} for the gauge fixing.
We thank Tatsumi Aoyama, Haozhao Liang, Shuangquan Zhang and Pengwei Zhao for helpful discussions.
 Y.L., H.T. and J.M.  were partially supported by the National Key R$\&$D Program of China (Contracts No. 2017YFE0116700 and No. 2018YFA0404400) and the
 National Natural Science Foundation of China (NSFC) under Grants No. 11935003, No. 11975031, No. 11875075, and No. 12070131001.
 This work was partially supported by JSPS Grant (No. JP18H05236, JP16H03978, JP19K03879, JP18H05407)
 and MOST-RIKEN Joint Project ``Ab initio investigation in nuclear physics'',
 ``Priority Issue on Post-K computer'' (Elucidation of the Fundamental Laws and Evolution of the Universe),
 ``Program for Promoting Researches on the Supercomputer Fugaku'' (Simulation for basic science: from fundamental laws of particles to creation of nuclei) and Joint Institute for Computational Fundamental Science (JICFuS).
\end{acknowledgments}


%



\clearpage
\begin{center}
\large{\bf{Supplemental Material}}
\end{center}

We present more details about the systematics related to the inelastic excited states and the derivative expansion of the non-local potential in this supplemental material.

\begin{figure}[htp]
  \centering
  \includegraphics[width=8.5cm]{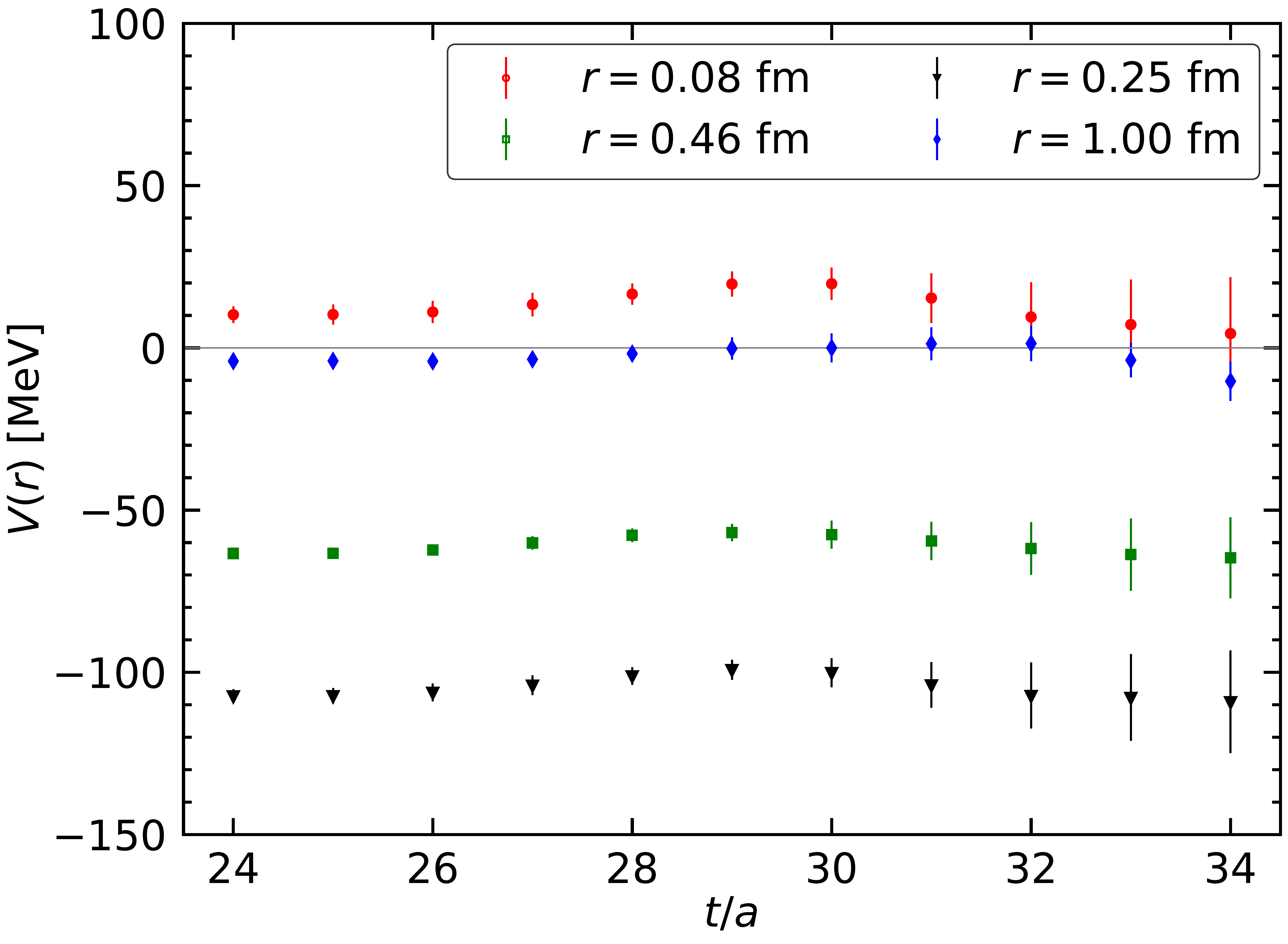}
  \caption{(Color online). The $t$-dependence of $\Omega_{ccc}\Omega_{ccc}$ potential $V(r)$ in the $^1S_0$ channel for several different values of the distance $r$.
} \label{Supp_Fig1}
\end{figure}

In Fig.~\ref{Supp_Fig1}, we plot the $t$-dependence of $\Omega_{ccc}\Omega_{ccc}$ potential $V(r)$ in the $^1S_0$ channel at several distances, $r=0.08$, $0.25$, $0.46$ and $1.00$ fm. We find that $V(r)$ at given $r$ varies slowly with $t$. This indicates the contributions from the inelastic excited states are small irrespective of the value of $r$. Note that the corresponding $t\in[2.0, 2.9]$ fm is sufficiently large compared to the scale relevant for the inelastic excited states, $\Lambda^{-1}_{\mathrm{QCD}}\sim0.7$ fm. 

The derivative expansion of the non-local potential up to the next-to-next-to-leading order (N$^2$LO) was studied in Ref.~\cite{Iritani2019PRDsupp}.
 It was  found that the leading order (LO) result is accurate enough for  heavy valence quarks at low energies.
 We expect the same to hold for  the $\Omega_{ccc}\Omega_{ccc}$ system in this study, and 
 leave  an explicit N$^2$LO analysis along the line with the above paper for future study. 
 In the main text,  we have estimated the systematic error from N$^2$LO by studying the $t$-dependence of the observables
since the truncation effect manifests itself through such $t$-dependence as  discussed  in the above mentioned  paper.

Yet another way to study the truncation effect in the derivative expansion was proposed in Ref.~\cite{Iritani2019Jhepsupp}. In the context of the present paper, it goes as follows.
  First, we construct a Hamiltonian $H$ in a finite box with the LO potential $V(r)$ obtained when $t/a=26$ as $H =-\nabla^2/m_{\Omega_{ccc}} + V(r)$.
   Eigenvalues and eigenfunctions of the LO Hamiltonian $H$ in the $A_1$ representation in a finite box
   are obtained from $H\psi_i=\epsilon_i\psi_i$. Then, 
   an improved two-baryon sink operator for a designated eigenfunction can be 
   constructed as $\sum_{\bm r}\psi^\dagger_i(\bm r)\Omega_{ccc}(\bm r, t)\Omega_{ccc}(\bm 0, t)$. 
Equivalently, we can define the generalized temporal correlation function as $R_{i}(t)\equiv\sum_{\bm r}\psi^\dagger_i(\bm r)R(\bm r, t)$, from which the effective energy shift for the $i$-th eigenstate can be defined as 
$\Delta E^{\mathrm{eff}}_i(t)= a^{-1} \log ({R_i(t)}/{R_i(t+1)})$. In Table.~\ref{tab1supp}, we show  $\Delta E^{\mathrm{eff}}_i$ fitting from the interval $t/a=24-28$ for the ground state ($i=0$) and the first excited state ($i=1$). Shown together are the energy shifts obtained by $\Delta E_i =2 \left(\sqrt{\epsilon_i\times m_{\Omega_{ccc}} +m_{\Omega_{ccc}}^2}-m_{\Omega_{ccc}} \right)$. 

\begin{table}[htbp]
\caption{The effective energy shift $\Delta E^{\mathrm{eff}}_i$ from the generalized temporal correlation function $R_{i}(t)$ and the corresponding energy shift $\Delta E_i$  from LO Hamiltonian $H$ for the ground state ($i=0$) and the first excited state ($i=1$) with statistical errors quoted in parentheses.}
\begin{tabular}{ccc}
  \hline\hline
   &~~~    $\Delta E^{\mathrm{eff}}_i~\mathrm{[MeV]}$  &~~~ $\Delta E_i~\mathrm{[MeV]}$ \\
  \hline
  $i=0$        &~~~ $-5.55(75)$ &~~~ $-5.87(85)$ \\
  $i=1$        &~~~ $0.52(23)$ &~~~ $0.53(25)$ \\
  \hline\hline
\end{tabular}
\label{tab1supp}
\end{table}

We find that $\Delta E^{\mathrm{eff}}_{0,1}$ from the generalized temporal correlation function $R_i(t)$  agree with the $\Delta E_{0,1}$  obtained from the LO Hamiltonian $H$ within the statistical errors.
 Although $R_i(t)$ utilizes the information of $H$ through eigenfunctions, the agreement of the energy shifts indicates 
 that higher order terms in the derivative expansion of the potential are not significant. 
 In other words, if the effect of higher order terms were large,  $\psi_i$ would be so different from the $i$-th eigenstate of the system that the effective energy shifts from $R_i(t)$ would  be distorted and do not  agree with those from $H$.

%


\end{document}